\documentclass[conference,a4paper]{IEEEtran}
\usepackage[utf8]{inputenc}
\usepackage[cmex10]{amsmath}
\usepackage{graphicx}
\usepackage{amssymb}

\usepackage[ruled,norelsize]{algorithm2e}
\makeatletter
\newcommand{\removelatexerror}{\let\@latex@error\@gobble}
\makeatother

\usepackage{algorithmicx}

\begin{document}
\title{Fast Channel Estimation for Millimetre Wave Wireless Systems Using Overlapped Beam Patterns}

\author{
    \IEEEauthorblockN{Matthew Kokshoorn, Peng Wang, Yonghui Li, and Branka Vucetic}
    \IEEEauthorblockA{        
        School of Electrical and Information Engineering, The University 	of Sydney, Australia \\
        Email: \{matthew.kokshoorn, peng.wang, yonghui.li,  branka.vucetic\}@sydney.edu.au }
}

\maketitle
\IEEEdisplaynontitleabstractindextext

\begin{abstract}
This paper is concerned with the channel estimation problem in millimetre wave (MMW) wireless systems with large antenna arrays. By exploiting the sparse nature of the MMW channel, we present an efficient estimation algorithm based on a novel overlapped beam pattern design. The performance of the algorithm is analyzed and an upper bound on the probability of channel estimation failure is derived. Results show that the algorithm can significantly reduce the number of required measurements in channel estimation (e.g., by 225\% when a single overlap is used) when compared to the existing channel estimation algorithm based on non-overlapped beam patterns. 
\end{abstract}

\section{Introduction}
Millimetre wave (MMW) communication has been shown to be a promising technique for next generation wireless systems due to the large expanse of available spectrum in the MMW frequency band, ranging from 30GHz to 300GHz. \cite{pi2011introduction}\cite{rappaport2013millimeter}. However, a critical challenge in exploiting the MMW frequency band is its severe signal propagation loss compared to that over conventional microwave frequencies \cite{rheath}\cite{zhang2010channel}. To compensate such a loss, large antenna arrays can be employed to achieve a high power gain. Fortunately, owing to the small wavelength of MMW signals, antennas can be packed into a small area at the transmitter and receiver \cite{hur2013millimeter}. 

\begin{figure}[!t]
\centering
\includegraphics[width=3.4in]{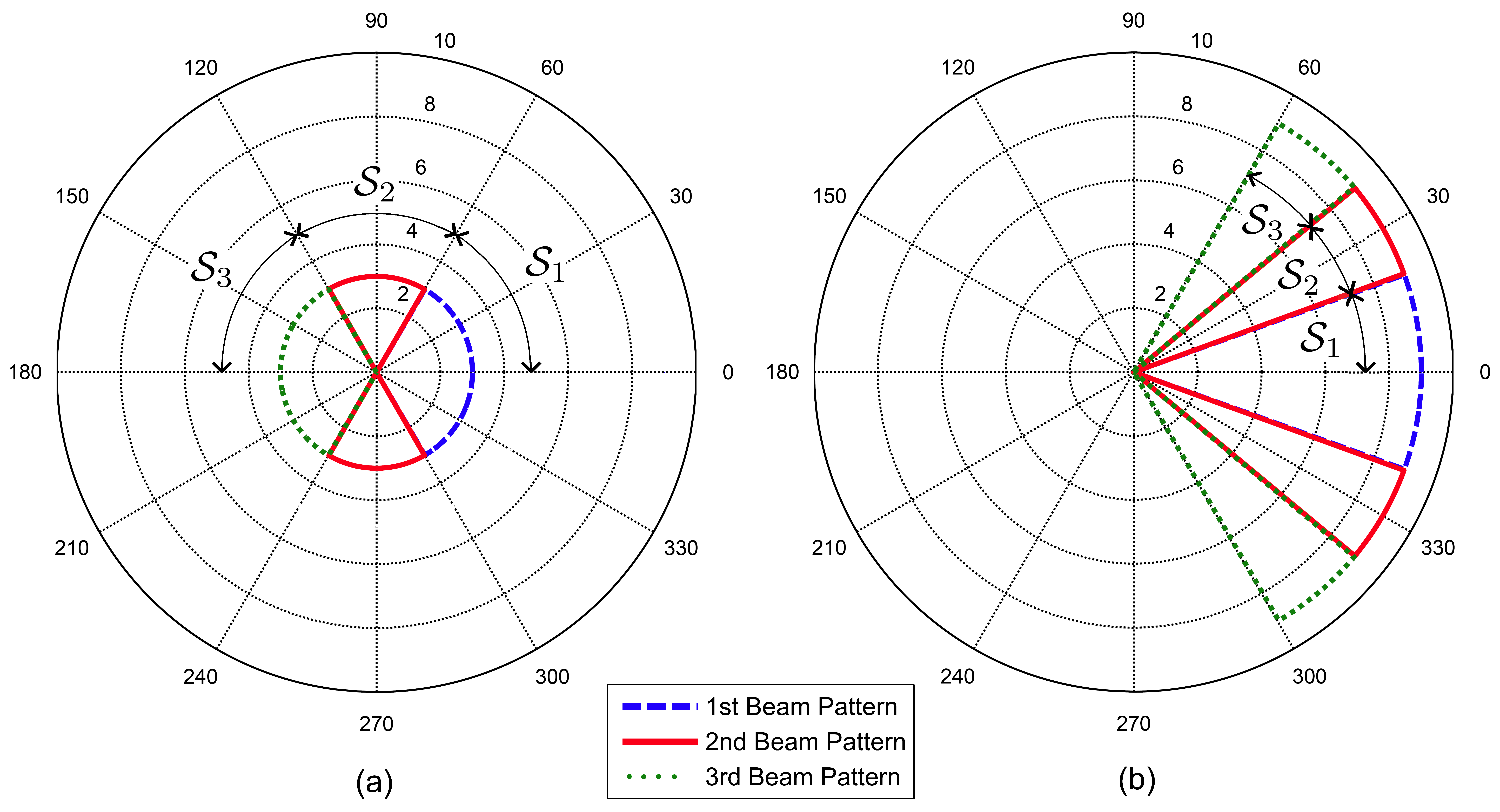}
\caption{Illustration of the beam patterns adopted in the first (a) and second (b) stages of the channel estimation algorithm of \cite{rheath} when $K = 3$. The three sub-ranges in the first stage are, $[0, \pi/3)$, $[\pi/3, 2\pi/3)$ and $[2\pi/3, \pi)$, respectively. By assuming that the possible AOAs/AODs are reduced to the sub-range $[0, \pi/3)$ in the first stage, this sub-range is further divided into $[0, \pi/9)$, $[\pi/9, 2\pi/9)$ and $[2\pi/9, \pi/3)$, respectively, in the second stage.}
\label{rheath_model}
\end{figure}

Channel state information (CSI) is essential for effective communication and precoder design in MMW systems. The use of large antenna arrays results in a large multiple-input multiple-output (MIMO) channel matrix. This makes the channel estimation a very challenging issue due to the large number of channel parameters to be estimated. On the other side, it was shown in recent channel measurements \cite{rappaportMeasure} that the MMW channel exhibits strong sparse propagation due to the high path loss. As a result, the MMW MIMO channel matrix can be represented as a sparse one after being converted into the angular space \cite{sayeed2002deconstructing}. Leveraging sparse geometric channel characteristics, \cite{rheath} developed a ``divide and conquer" type multi-stage algorithm for estimating sparse MMW channels. As shown in \figurename  \ref{rheath_model}, in each stage of this algorithm, the possible ranges of angles of departure (AODs) and angles of arrival (AOAs) are both divided into $K$ non-overlapped angular sub-ranges. Correspondingly, $K$ non-overlapped beam patterns are designed at both the transmitter and receiver such that each transmit (receive) beam pattern exactly covers one AOD (AOA) angular sub-range. The channel estimation carried out in each stage consists of $K^2$ time slots. In each time slot, the pilot signal is transmitted using one of the $K$ beam patterns at the transmitter, and then received by one of the $K$ beam patterns at the receiver, the corresponding channel output for this transmit and receive beam pattern can then be obtained. These $K^2$ time slots span all the combinations of transmit-receive beam patterns. By comparing the magnitudes of the corresponding $K^2$ channel outputs, the transmit/receive sub-ranges that the AOD/AOA  most likely  belong to are determined. Afterwards, the algorithm will limit the
estimation to the angular sub-range identified at each link end in the previous stage and further divide it into $K$ sub-ranges for the channel estimation in next stage. This process continues until the smallest beam width resolution is reached. It is shown in \cite{rheath} that the algorithm requires estimation time proportional to $K^2\lceil \text{log}_K (\text{max}(N_t,N_r))\rceil$ per path where $N_t$ and $N_r$ are, respectively, the numbers of transmit and receive antennas. In rapidly varying channels, such a channel estimation algorithm may not be quick enough to track the fast channel variations. Therefore it is desirable to develop a fast and efficient algorithm to further reduce the estimation time.
 
In this paper, we propose a fast channel estimation algorithm by designing a set of novel overlapped beam patterns that can significantly reduce the number of measurement time slots required for channel estimation. The overlap between the beam patterns increases the amount of information carried by the channel outputs, and thus a path can then be identified by using a combination of multiple channel outputs, instead of using just a single one. In this way, the proposed algorithm can reduce the number of time slots required for channel estimation and thus speed up the estimation process. We also develop a minimum mean squared error (MMSE) channel estimator to estimate the channel coefficient by optimally combining the selected measurements in all stages. The performance of the proposed algorithm is analyzed and an upper bound on the probability of channel estimation failure is derived. The analysis is then validated by Monte Carlo simulations.

\section{System Model}
Consider a MMW MIMO system where both the transmitter and receiver are equipped with a half-wavelength spacing uniform linear antenna array (ULA). For simplicity, in this paper we assume the same number of antennas, denoted by $N$, at both the transmitter and receiver. The proposed algorithm can be easily extended to a general asymetric system. Assume that there is a single-path channel between the transmitter and receiver with AOD, $\phi \in [0,\pi)$, and AOA, $\theta \in [0,\pi)$. Then the corresponding channel matrix can be represented as\footnote{Note that the use of ULA results in no distinguishable difference between AOAs $\theta$ and $-\theta$ or between AODs $\phi$ and $-\phi$. Hence, only AODs and AOAs in the range $[0,\pi)$ need to be considered.} 

\begin{equation}
\label{H}
\boldsymbol{H} = \alpha \boldsymbol{a}_{r}(\theta) \boldsymbol{a}_{t}^{H}(\phi)
\end{equation} 

\noindent    
where $\alpha$ is the channel fading coefficient, $(\cdot)^H$ denotes the conjugate transpose operation and $\boldsymbol{a}_r(\theta)$ and $\boldsymbol{a}_t(\phi)$ are, respectively, the receive and transmit spatial signatures of the single path. They are defined as $\boldsymbol{a}_r(\theta)= \boldsymbol{u}(\theta)$ and $\boldsymbol{a}_t(\phi)= \boldsymbol{u}(\phi)$, respectively, where 

\begin{equation}
\label{u_n}
\boldsymbol{u}(\epsilon) \triangleq \frac{1}{\sqrt{N}} [1,e^{j \pi \text{sin}(\epsilon)},\cdots,e^{j\pi(N-1)\text{sin}(\epsilon)}]^T
\end{equation}

\noindent
and $(\cdot)^T$ is the transpose operation. Based on (\ref{H}), the overall channel state information only includes three parameters, i.e., the AOA $\theta$, the AOD $\phi$, and the fading coefficient $\alpha$. We assume that both $\theta$ and $\phi$ can only take some discrete values from the set $\{0,\frac{\pi}{N},\cdots,\frac{\pi(N-1)}{N} \}$. We aim to find an efficient way to estimate these three parameters. 

We assume that both the transmitter and receiver are equipped with a limited number of RF chains. Following \cite{rheath}, we further assume that these RF chains, at one end, can only be jointly adopted to form a single beam pattern, indicating that only one pilot signal can be transmitted and received at one time. Define $\boldsymbol{f}$ and $\boldsymbol{w}$  $(||f||_2 = ||w||_2 = 1)$, respectively, as the $N \times 1$ unit beamforming vector at the transmitter and $N \times 1$ unit combining vector at the receiver. The corresponding channel output can be represented as

\begin{equation}
\label{y}
y = \sqrt{p} \boldsymbol{w}^H \boldsymbol{H} \boldsymbol{f} x + \boldsymbol{w}^{H} \boldsymbol{n}
\end{equation}

\noindent
where $x$ is the transmitted pilot signal with unit power, $p$ is the transmit power and $\boldsymbol{n}$ is a length-$N$ vector of independent and identically distributed (i.i.d.) complex additive white Gaussian noise (AWGN) samples with mean zero and variance $N_0$. The key challenge here is how to design a sequence of $\boldsymbol{f}$ and $\boldsymbol{w}$ in such a way that the channel parameters can be quickly estimated without an exhaustive search of all beam angles. 

\section{Proposed Channel Estimation with Overlapped Beam Patterns}
In this section, we propose a new set of beam patterns that are overlapped with one another, based on which a fast channel estimation algorithm is designed to accurately retrieve the channel state information.

\begin{figure}[!t]
\centering
\includegraphics[width=3.4in]{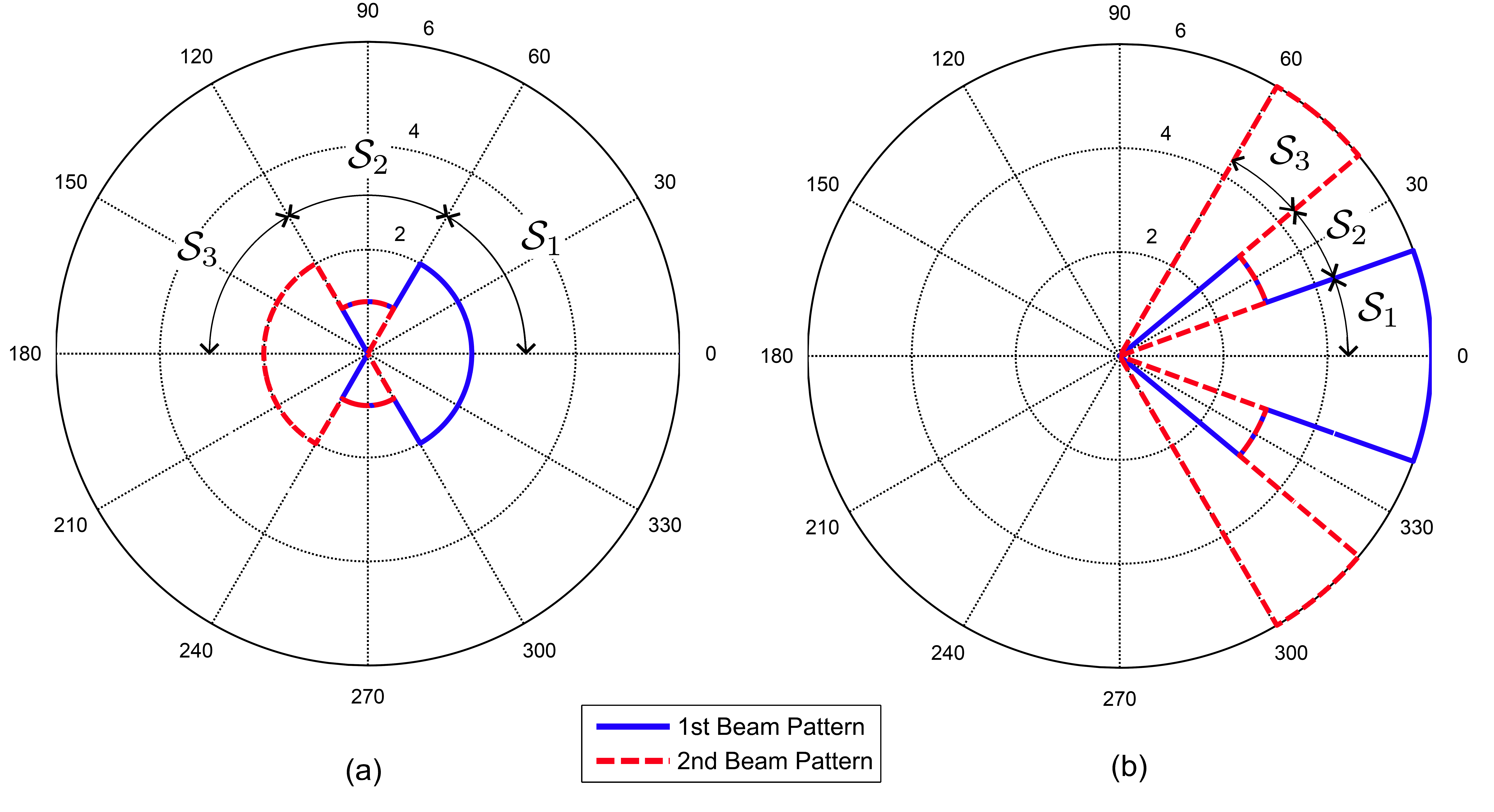}
\caption{Illustration of the overlapped beam patterns adopted in the first (a) and second (b) stages of the proposed algorithm when $K = 3$. }
\label{US_K3}
\end{figure}         

\subsection{An Example}
To better understand our proposed overlapped beam pattern design, here we first consider an example. Following \figurename \ref{rheath_model}, we still divide the  AOD/AOA angular spaces into $K = 3$  sub-ranges in the first stage, denoted by $\mathcal{S}_{1} = [0, \pi/3)$, $\mathcal{S}_{2} = [\pi/3, 2\pi/3)$ and $\mathcal{S}_{3} = [2\pi/3, \pi)$, respectively. Instead of using 3 beam patterns to cover them at each link end as in \figurename \ref{rheath_model}(a), we aim to use only $M = 2$ overlapped beam patterns to achieve this. \figurename \ref{US_K3}(a) illustrates our designed beam patterns in the first stage. We can see that the first and second beam patterns cover $\mathcal{S}_1$, $\mathcal{S}_2$ and $\mathcal{S}_2$, $\mathcal{S}_3$, respectively, and are overlapped in the whole range of $\mathcal{S}_2$. It is also seen that each beam pattern can have different amplitudes in different sub-ranges. The beam patterns can be described using a matrix, referred to as the beam pattern description matrix, as

\begin{equation}
\label{B_generic}
    \boldsymbol{B} =
    \left[\begin{array}{ccc}
    b_{1,1}&b_{1,2}&b_{1,3}\\
    b_{2,1}&b_{2,2}&b_{2,3}
    \end{array}\right]
\end{equation}

\noindent
where $b_{m,k}$ denotes the amplitude of the $m^{th}$ beam pattern in sub-range $S_{k}$. For the beam patterns in \figurename \ref{US_K3}, we have $b_{1,3}=b_{2,1}=0$, as beam patterns $1$ and $2$ do not cover, respectively, the $3^{rd}$ and $1^{st}$ sub-ranges. Due to the symmetry between the two beam patterns, we further have $b_{1,1} = b_{2,3} = \beta_1$ and $b_{1,2} = b_{2,2} = \beta_2$.
Furthermore, it is always desirable that the same quantity of signal power is transmitted/received via different sub-ranges during one stage. In order to achieve this, we should normalize each column of $\boldsymbol{B}$. This leads to $\beta_1 = 1$ and $\beta_2 = \frac{1}{\sqrt{2}}$, and the matrix $\boldsymbol{B}$ reduces to

\begin{equation}
\label{B_zero}
    \boldsymbol{B} =
    \left[\begin{array}{ccc}
    \beta_1&\beta_2&0\\
    0&\beta_2&\beta_1
    \end{array}\right]=
    \left[\begin{array}{ccc}
    1&\frac{1}{\sqrt{2}}&0\\
    0&\frac{1}{\sqrt{2}}&1\\
    \end{array}\right].
\end{equation} 

To generate the beam patterns illustrated in \figurename \ref{US_K3}(a), the beamforming/combining vectors should be designed as follows. Denote by $\boldsymbol{f}_m$ and $\boldsymbol{w}_m$, respectively, the beamforming vector and combining vector corresponding to the $m^{th}$ beam pattern in \figurename \ref{US_K3}(a). We then have

\begin{equation}
\label{range_example}
\boldsymbol{u}^H(\epsilon)\boldsymbol{f}_m \! =\! \boldsymbol{u}^H(\epsilon) \boldsymbol{w}_m \!= \! C b_{m,k},\! \text{if \!$ \exists \; k \in \! \{ 1,2 ,3 \},\epsilon \! \in \! \mathcal{S}_{k}  $},
\end{equation}

\noindent 
where $\boldsymbol{u}(\epsilon)$ has been defined in (\ref{u_n}) and $C$ is a scalar constant that ensures $||\boldsymbol{f}_m||_2=||\boldsymbol{w}_m||_2=1$. Equation (\ref{range_example}) can be expressed in a matrix form as

\begin{equation}
\label{matrix_eq}
\boldsymbol{U}^H \boldsymbol{f}_m=\boldsymbol{U}^H \boldsymbol{w}_m= \left[\begin{array}{ccc}
    C{b}_{m,1} \boldsymbol{1}_{|S_{1}|} \\
    C{b}_{m,2} \boldsymbol{1}_{|S_{2}|} \\
    C{b}_{m,3} \boldsymbol{1}_{|S_{3}|} 
    \end{array}\right]
     \triangleq \boldsymbol{g}_m
\end{equation}

\noindent 
where $\boldsymbol{U} = \Big[ \boldsymbol{u}(0),\boldsymbol{u} \Big( \frac{\pi}{N} \Big),\cdots,\boldsymbol{u} \Big( \frac{\pi(N-1)}{N} \Big) \Big]$, $\boldsymbol{1}_{|S_{k}|}$ is a length-$|S_{k}|$ all-one column vector and $|S_{k}|$ returns the size of set $S_{k}$. Therefore $\boldsymbol{f}_m$ and $\boldsymbol{w}_m$ can be designed as 

\begin{equation}
\label{f_m}
\boldsymbol{f}_m = \boldsymbol{w}_m  = (\boldsymbol{U^H})^{-1} \boldsymbol{g}_m.
\end{equation}
  
We now perform channel estimation in the first stage using the above designed beamforming/combining vectors $\{\boldsymbol{f}_m\}$ and $\{\boldsymbol{w}_m\}$. In each time slot, we select one beamforming vector $\boldsymbol{f}_m$ at the transmitter and one combining vector $\boldsymbol{w}_n$  at the receiver to transmit/receive the pilot signal $x$. The corresponding channel output, denoted by $y_{n,m}$, is given by

\begin{equation}
\label{y_n_m}
y_{n,m} = \sqrt{p} \boldsymbol{w}_n^H \boldsymbol{H} \boldsymbol{f}_m x + q_{n,m}
\end{equation}

\noindent
where $q_{n,m}$ is the corresponding noise term in $y_{n,m}$. By using $M^2$ time slots to span all the combinations of $\{\boldsymbol{f}_m\}$ and $\{\boldsymbol{w}_n\}$, we generate a total number of $M^2$ channel outputs that can be represented in a matrix form as

\begin{align}
\label{Y}
\boldsymbol{Y} =& \sqrt{p} \boldsymbol{W}^H \boldsymbol{H}\boldsymbol{F} x +\boldsymbol{Q} \nonumber \\
 =& \alpha \sqrt{p} (\boldsymbol{u}^H(\theta)\boldsymbol{W})^H (\boldsymbol{u}^H(\phi)\boldsymbol{F})^H x +\boldsymbol{Q} 
\end{align}

\noindent
where $\boldsymbol{F}=[\boldsymbol{f}_1,\cdots,\boldsymbol{f}_{M}]$, $\boldsymbol{W}=[\boldsymbol{w}_1,\cdots,\boldsymbol{w}_{M}]$ and $\boldsymbol{Q} = \{q_{m,n}\}$ is an $M\times M$ matrix.

Since we only have $M = 2$ beam patterns at each link end, a total number of $M^2 = 4$  channel outputs, as described in (\ref{Y}), can be obtained. However, there are in total $K^2=9$ transmit-receive sub-range combinations denoted by $\{(k_t, k_r)| k_t, k_r = 1 ,2, 3\}$ where $k_t$ and $k_r$ are respectively, the transmit and receive sub-range indices. To accurately extract the AOA/AOD information from $\boldsymbol{Y}$, let us assume that AOD, $\phi \in \mathcal{S}_{k_t}$ and AOA, $\theta \in \mathcal{S}_{k_r}$, without loss of generality. In this case, by recalling (\ref{range_example}) we have 

    \begin{eqnarray}
    \label{simplify}
	\boldsymbol{u}^H(\phi) \boldsymbol{f}_m = C b_{m, k_t} \text{ and } \boldsymbol{u}^H(\theta) \boldsymbol{w}_m = C b_{m, k_r}.
    \end{eqnarray}

\noindent
Hence (\ref{Y}) can be re-written as

    \begin{eqnarray}
    \label{Y_3}
    \boldsymbol{Y} = \alpha \sqrt{p} C^2 \boldsymbol{b}_{k_r} \boldsymbol{b}_{k_t}^T   x  + \boldsymbol{Q}
    \end{eqnarray}

\noindent
where $\boldsymbol{b}_{i}$ denotes the $i^{th}$ column of $\boldsymbol{B}$ in (\ref{B_zero}). By further vectorizing (\ref{Y_3}), we obtain 

   \begin{eqnarray}
    \label{Y_vec}
    \text{vec}(\boldsymbol{Y}) &=& \alpha \sqrt{p} C^2 \text{vec}(\boldsymbol{b}_{k_r} \boldsymbol{b}_{k_t}^T)   x  +\text{vec}( \boldsymbol{Q}) \nonumber \\
     &=& \alpha \sqrt{p} C^2 (\boldsymbol{b}_{k_t} \otimes \boldsymbol{b}_{k_r})x     + \text{vec}(\boldsymbol{Q})   
    \end{eqnarray}
    
\noindent    
where $\boldsymbol{b}_{k_t} \otimes \boldsymbol{b}_{k_r}$ denotes the Kronecker product between $\boldsymbol{b}_{k_t}$ and $\boldsymbol{b}_{k_r}$. It is worth noting that the vector $\boldsymbol{b}_{k_t} \otimes \boldsymbol{b}_{k_r}$ has a unit norm as $\boldsymbol{b}_{k_t}$ and $\boldsymbol{b}_{k_r}$ are both unit vectors.

To detect the pilot signal $x$ from $\text{vec}(\boldsymbol{Y})$, here we adopt the maximum ratio combining (MRC) principle and correlate $\text{vec}(\boldsymbol{Y})$ using the unit vector $\boldsymbol{b}_{k_t} \otimes \boldsymbol{b}_{k_r}$. The resultant correlation output, referred to as a channel measurement, is given by

   \begin{eqnarray}
    \label{r}
     r_{{k}_r,{k}_t}  &=&  (\boldsymbol{b}_{{k}_t} \otimes \boldsymbol{b}_{{k}_r})^T  \text{vec}(\boldsymbol{Y}^{(s)}) = \boldsymbol{b}_{{k}_r}^T \boldsymbol{Y} \boldsymbol{b}_{{k}_t}.
    \end{eqnarray}
    
\noindent
Note that the above discussions (\ref{simplify})-(\ref{r}) are based on the assumption of $\phi \in \mathcal{S}_{k_t}$ and $\theta \in \mathcal{S}_{k_r}$. If this assumption is correct, (\ref{r}) reduces to 

\begin{eqnarray}
\label{r_corr}
r_{k_r, k_t} = \alpha \sqrt{p} C^2 x + \boldsymbol{b}_{k_r}^T \boldsymbol{Q} \boldsymbol{b}_{k_t},
\end{eqnarray}

\noindent
yielding an SNR of $\frac{ |\alpha|^2 p C^4}{N_0}$ for the pilot signal $x$. Otherwise if the assumption is incorrect, i.e., $\phi \in \mathcal{S}_{k_t'} \text{ and } \theta \in \mathcal{S}_{k_r'}$ for some $k_t' \neq k_t \text{ and/or } k_r' \neq k_r$, (\ref{r}) reduces to 

\begin{eqnarray}
\label{r_incorr}
r_{k_r, k_t} = \alpha \sqrt{p} C^2 \boldsymbol{b}_{{k}_r}^T \boldsymbol{b}_{k_r'} \boldsymbol{b}_{k_t'}^T \boldsymbol{b}_{{k}_t}  x + \boldsymbol{b}_{k_r}^T \boldsymbol{Q} \boldsymbol{b}_{k_t} 
\end{eqnarray}

\noindent
in which the SNR is given by $(\boldsymbol{b}_{k_r}^T\boldsymbol{b}_{k_r'} \boldsymbol{b}_{k_t'}^T \boldsymbol{b}_{{k}_t})^2 \frac{ |\alpha|^2 p C^4 }{N_0} $ that is always no larger than half of the SNR in (\ref{r_corr}). By considering all possible values of $k_r$ and $k_t$, we can obtain a total number of $K^2= 9$ channel measurements $\{r_{k_r, k_t}\}$ that can form a $K \times K$ matrix $\boldsymbol{R}$, i.e.,   

\begin{equation}
\label{R}
\boldsymbol{R} = \boldsymbol{B}^T \boldsymbol{Y} \boldsymbol{B}.
\end{equation}

\noindent
Finally, by finding,

\begin{equation}
\label{US_decision}
(\hat{k}_r,\hat{k}_t) =  \underset{{k}_r,{k}_t=1,\cdots,K}{\operatorname{argmax}} |{r}_{{k}_r,{k}_t}|,
\end{equation} 

\noindent
we can reduce the ranges of possible AOAs and AODs to, respectively, the $\hat{k}_t^{th}$ and $\hat{k}_r^{th}$ transmit and receive angular sub-ranges. Each of these two sub-ranges will be further divided into another $K$ sub-ranges for the channel estimation in the next stage.

\subsection{Extension to General Cases} 

In general, for a single-path MMW channel, our proposed channel estimation algorithm also works in a similar multi-stage manner as that in \cite{rheath}. In each stage, denoted $s$, we divide the possible AOA angular space into $K=2^M-1$ (where $M$ is an integer) non-overlapped sub-ranges $\mathcal{S}^{(s)}_{r,1}, \mathcal{S}^{(s)}_{r,2}, \dots, \mathcal{S}^{(s)}_{r,K}$ and divide similarly the possible AOD angular space into $\mathcal{S}^{(s)}_{t,1}, \mathcal{S}^{(s)}_{t,2}, \cdots, \mathcal{S}^{(s)}_{t,K}$. Then only $M$ overlapped beam patterns will be designed at each end to cover these $K$ sub-ranges. The designed $M$ beam patterns are characterized by an $M \times K$ beam pattern description matrix $\boldsymbol{B}$, which consists of the normalized version of all the $K = 2^M-1$ non-zero length-$M$ binary vectors as its columns. Besides the example of $\boldsymbol{B}$ with $M = 2$ and $K = 3$ in (\ref{B_zero}), another example with $M = 3$ and $K = 7$ is given by\footnote{Note that the columns of $\boldsymbol{B}$ can be arbitrarily permuted, which does not affect the performance of our proposed channel estimation algorithm. However, a better column-permutation of $\boldsymbol{B}$, e.g., a gray-coded based permutation, may facilitate the realization of the corresponding ${\boldsymbol{f}_m}$ and ${\boldsymbol{w}_m}$ when the hardware constraints are considered.}

\begin{equation}
\label{K7}
    \boldsymbol{B}=\left[\begin{array}{ccccccc}
    1\;&\frac{1}{\sqrt{2}}\;&\frac{1}{\sqrt{3}}\;&\frac{1}{\sqrt{2}}\;&0\;&0\;&0\\
    0\;&0\;&\frac{1}{\sqrt{3}}\;&\frac{1}{\sqrt{2}}\;&1\;&\frac{1}{\sqrt{2}}\;&0\\
    0\;&\frac{1}{\sqrt{2}}\;&\frac{1}{\sqrt{3}}\;&0\;&0\;&\frac{1}{\sqrt{2}}\;&1
    \end{array}\right].
\end{equation}    

Given $\boldsymbol{B}$, we can then generate both the beamforming vectors $\{\boldsymbol{f}^{(s)}_{m}\}$ and combining vectors $\{\boldsymbol{w}^{(s)}_{n}\}$ in the same way as (\ref{f_m}). For example, to generate ${\boldsymbol{f}^{(s)}_{m}}$, the corresponding vector $\boldsymbol{g}_{m}$ in (\ref{f_m}), which is redefined as $\boldsymbol{g}_{m,t}^{(s)}$ for rigorousness, should be designed such that its $i^{th}$ entry, denoted by $g_{m,t}^{(s)}(i)$, satisfies

    \begin{align}
    \label{desired_output_gen}
    &g_{m,t}^{(s)}(i)\!= \begin{cases}
    C_s b_{m,k},\!&\text{\!if $\frac{\pi i}{N} \in {\mathcal{S}_{t,k_t}^{(s)}, }\;\exists \;k_t \in \{1,\cdots,K \} ;$\!}  \\
    $0$,\!&\text{\!if $\frac{\pi i}{N} \notin {\mathcal{S}_{t,k_t}^{(s)}, } \;\forall \; k_t \in \{1,\cdots,K \} $\!} \!
    \end{cases}
    \end{align}

\begin{figure}[!t]
\removelatexerror
\begin{algorithm}[H]
\label{alg1}

\caption{Single-path channel estimation algorithm for MMW channels.}

{\fontsize{9}{9}\selectfont
$\mathbf{Initialization:}$ $\mathcal{S}_{t,k}^{(1)},\mathcal{S}_{r,k}^{(1)} \;\forall \;k=1,\cdots,K$ \\
						 
   \For( \emph{}){$s\leq S $}
   {
   	   Calculate $\boldsymbol{F}^{(s)}$ based on $\mathcal{S}_{t,k}^{(s)}\;\forall \;k=1,\cdots,K$ and $\boldsymbol{B}$ \\
   	   Calculate $\boldsymbol{W}^{(s)}$ based on $\mathcal{S}_{r,k}^{(s)}\;\forall \;k=1,\cdots,K$ and $\boldsymbol{B}$ 
   	  
       \For( \emph{}){$m=1$ to $M$}
        {   
          Transmitter transmits using $\boldsymbol{f}^{(s)}_{m}$ \\
          \For( \emph{}){$n=1$ to $M$}
            {         
                Receiver measures using $\boldsymbol{w}^{(s)}_{n}$
            }    
            
        }
            After $M^2$ measurements:
                 $\boldsymbol{Y}^{(s)}= \sqrt{p_{s}} (\boldsymbol{{W}^{(s)}})^H \boldsymbol{H} \boldsymbol{F}^{(s)}x  + \boldsymbol{Q}$    
          
        $\boldsymbol{R}^{(s)}= \boldsymbol{B}^T\boldsymbol{Y}^{(s)}\boldsymbol{B}$
 
 		$(\hat{k}^{(s)}_r,\hat{k}^{(s)}_t) = \underset{{{k}_r,{k}_t=1,\cdots,K}}{\operatorname{argmax}} |{r}^{(s)}_{{k}_r,{k}_r}|$
        
   }
   
   $\hat{\phi}=\frac{\pi}{N} \sum\limits_{s=1}^S (\hat{k}^{(s)}_t-1)K^{S-s}  ,\;\hat{\theta}= \frac{\pi}{N} \sum\limits_{s=1}^S (\hat{k}^{(s)}_r-1)K^{S-s}$
   
      $\hat{\alpha}= P_R \boldsymbol{h}^H(\boldsymbol{h} P_R \boldsymbol{h}^H + N_0 \boldsymbol{I}_S)^{-1} \boldsymbol{\boldsymbol{r}}$
   }
\end{algorithm}
\end{figure}

\noindent
where $C_s$ is a scalar constant that satisfies $||\boldsymbol{f}_m^{(s)}||_2=1$. Physically, $g_{m,t}^{(s)}(i)$ describes the desired beam pattern amplitude at angle, $\frac{\pi i}{N}$ when ${\boldsymbol{f}^{(s)}_{m}}$ is used. Each combining vector ${\boldsymbol{w}^{(s)}_{n}}$ can be designed in the same way.

The channel output on the $s^{th}$ estimation stage can then be obtained via $M^2$ time slots by

\begin{align}
\label{Y_s}
\boldsymbol{Y}^{(s)} = \sqrt{p_s} (\boldsymbol{W}^{(s)})^H \boldsymbol{H}\boldsymbol{F}^{(s)}  x +\boldsymbol{Q}^{(s)}
\end{align}

\noindent
where $p_s$ denotes the transmit power of the pilot signal in the $s^{th}$ stage. Similar to that in \cite{rheath}, we prefer that all the stages have an equal probability of failure, indicating that we can allocate power among stages inverse proportionally to the beamforming gains of these beam patterns, i.e., 

    \begin{equation}
    \label{p_s}
    {p_{s}}=\frac{{P_T}}{C_{s}^4} \; \forall \; s=1,2,\cdots
    \end{equation}
    
\noindent
where $P_T$ is a constant. Similarly to (\ref{R}) we convert $\boldsymbol{Y}^{(s)}$ into the measurement matrix

    \begin{equation}
    \label{R_s}
 	\boldsymbol{R}^{(s)} = \boldsymbol{B}^T \boldsymbol{Y}^{(s)}\boldsymbol{B} .
    \end{equation}

\noindent   
The most likely AOD/AOA combination is determined using

\begin{equation}
\label{US_decision_general}
(\hat{k}_r^{(s)},\hat{k}_t^{(s)}) =  \underset{{k}_r,{k}_t=1,\cdots,K}{\operatorname{argmax}} |{r}_{{k}_r,{k}_t}^{(s)}|
\end{equation} 

\noindent
where ${r}_{{k}_r,{k}_t}^{(s)}$ is the $({{k}_r^{(s)},{k}_t^{(s)}} )^{th}$ entry of $\boldsymbol{R}^{(s)}$. The selected sub-ranges, $\mathcal{S}_{t,\hat{k}_t^{(s)}}^{(s)}$ and $\mathcal{S}_{r,\hat{k}_r^{(s)}}^{(s)}$ are used for the channel estimation on the next stage. 

Once $S=\lceil \text{log}_KN \rceil$ estimation stages of Algorithm \ref{alg1} have been completed, the minimum angle resolution $\frac{\pi}{N}$ is reached and the fading co-efficient, $\alpha$, can be estimated. \cite{rheath} estimated ${\alpha}$ based on only the measurement of the final stage. To improve the estimate accuracy, in this paper we propose to estimate ${\alpha}$ by using the selected measurement in all stages of the algorithm. Denote by $r_{\hat{k}_r^{(s)}, \hat{k}_t^{(s)}}^{(s)}$ the selected measurement from $\boldsymbol{R}^{(s)}$ in the $s^{th}$ stage. Define

    \begin{equation}
    \label{r_bold_vec}
	\boldsymbol{r} = [r_{\hat{k}_r^{(1)}, \hat{k}_t^{(1)}}^{(1)}, r_{\hat{k}_r^{(2)}, \hat{k}_t^{(2)}}^{(2)},\cdots,r_{\hat{k}_r^{(S)}, \hat{k}_t^{(S)}}^{(S)}]^{T}.   
    \end{equation}
    
\noindent
Then from (\ref{r_corr}) and (\ref{p_s}), we can see that, provided that the selection is correct in each stage, (\ref{r_bold_vec}) can be rewritten as 

    \begin{equation}
    \label{r_bold}
	\boldsymbol{r} = \sqrt{P_T} x \cdot \boldsymbol{1}_S \alpha + \boldsymbol{n}     
    \end{equation}

\noindent
where $\boldsymbol{n}$ is the $S\times 1$ vector of corresponding noise terms. Following the MMSE principle, we can then estimate the fading coefficient $\alpha$ as

\begin{equation}
\label{alpha_est} 
   \hat{\alpha}= \text{Var}[\alpha] \sqrt{P_T} x^H \boldsymbol{1}_S^H(\text{Var}[\alpha] {P_T} \cdot \boldsymbol{1}_S\boldsymbol{1}_S^H + N_0 \boldsymbol{I}_S)^{-1} \boldsymbol{\boldsymbol{r}}
\end{equation}

\noindent
where $\text{Var}[\alpha]$ is the variance of $\alpha$, and $\boldsymbol{I}_S$ is an $S\times S$ identity matrix.

It can be seen that, compared with the channel estimation algorithm in \cite{rheath} with the same value of $K$, our proposed algorithm also requires $S=\lceil \text{log}_K N \rceil$ stages, but the number of time slots required in each stage reduces to $M^2=\text{log}_2^2(K+1)$, instead of $K^2$. In general, this yields a $\frac{K^2}{\text{log}_2^2(K+1)}$ reduction in measurement time slots. For the example of $K=3$ discussed earlier, a 225\% reduction of measurement time slots can be achieved. Table 1 tabulates the total numbers of time slots required of the proposed algorithm and the algorithm in \cite{rheath} over a range of antenna array sizes. It is seen a significant reduction in required time slots can be achieved with the proposed algorithm with both $K=3$ and $K=7$.

\section{Performance Analysis}
\label{sec:Probability}
In this section, we analyze the performance of our proposed channel estimation algorithm. We say that the channel estimation fails if the selected transmitter-receiver sub-range pair at the final stage $S$ does not contain the correct AOD/AOA of the channel path. The corresponding probability of channel estimation failure (PCEF) can be expressed as  

    \begin{equation}
    \label{fail_cond}
    {P_{fail}}(S)= \text{Pr}\Big( \phi \notin \mathcal{S}_{t,\hat{k}_t^{(S)}}^{(S)} \text{ or } \theta \notin \mathcal{S}^{(S)}_{r,\hat{k}_r^{(S)}} \Big).
    \end{equation}

\noindent
Define $P_{fail}(s|s-1)$ as the probability that the channel estimation succeeds at stage $s-1$ but fails at stage $s$, i.e.,

    \begin{align}
    	P_{fail}  (s|s-1)  \triangleq  \text{Pr} \Big( & \phi \notin \mathcal{S}_{t,\hat{k}_t^{(s)}}^{(s)} \text{ or } \theta \notin \mathcal{S}^{(s)}_{r,\hat{k}_r^{(s)}} \nonumber \\
    	&\Big| \phi \in \mathcal{S}_{t,\hat{k}_t^{(s-1)}}^{(s-1)} \text{ and } \theta \in \mathcal{S}^{(s-1)}_{r,\hat{k}_r^{(s-1)}} \Big). 
    \end{align}
    
\noindent
Then (\ref{fail_cond}) can be upper bounded by

\begin{table}[!t]
\renewcommand{\arraystretch}{1.3}
\caption{Comparison of the number of time slots required for the proposed overlapped algorithm and the non-overlapped algorithm in \cite{rheath}.}
\label{table_1}

\centering
\begin{tabular}{|c||c||c|}
\hline 
\hspace{18.15 mm} & \hspace{8.1 mm} $K$=3 \hspace{6.75 mm} & \hspace{9.2 mm} $K$=7 \hspace{11.2 mm} \\  

\end{tabular}

\begin{tabular}{|c||c|c|c|c||c|c|c|c|}
\hline 
$N$ & 3 & 9 & 27 & 81 & 7 & 49 & 343 & 2401 \\
\hline 
\hline 
Overlapped & 4 & 8 & 12 & 16 & 9 & 18 & 27 & 36 \\
\hline 
Non-overlapped & 9 & 18 & 27 & 36 & 49 & 98 & 147 & 196 \\
\hline 
\end{tabular}
\end{table}

    \begin{equation}
    \label{fail_cond_full}
    P_{fail}(S) = 1-\prod_{s=1}^{S} (1-{P_{fail}(s|s-1)}) \leq \sum_{s=1}^{S} P_{fail}(s|s-1), \\            
    \end{equation}
    
\noindent    
in which each additive term $P_{fail}(s|s-1)$ can be further elaborated as follows. Denote by $k_t'^{(s)}$ and $k_r'^{(s)}$, respectively, the indices of the sub-ranges where the AOD and AOA of the single path falls in at the $s^{th}$ stage, i.e., $\phi \in \mathcal{S}_{t,{k}_t'^{(s)}}^{(s)}$ and $\theta \in \mathcal{S}^{(s)}_{r,{k}_r'^{(s)}}$. Then physically, $P_{fail}(s|s-1)$ corresponds to the event that the correct measurement $r_{{k}_r'^{(s)},{k}_t'^{(s)}}^{(s)}$ does not have the largest magnitude among all entries of $\boldsymbol{R}^{(s)}$. Mathematically, we have

  \begin{align}
  \label{P_fail_eq_2}
  		&P_{fail}(s|s-1) \nonumber \\
  		&= \frac{1}{K^2} {\sum_{{{k}_r'^{(s)} ,{k}_t'^{(s)} =1}}^{K}}  \text{Pr} \Big( \underset{{k}_r,{k}_t\; = \;1,\cdots,K}{\operatorname{max}} |r_{{k}_r,{k}_t}^{(s)}| > |r_{{k}_r'^{(s)},{k}_t'^{(s)}}^{(s)}| \Big) \nonumber \\
     & \leq  \frac{1}{K^2} {\sum_{{{k}_r'^{(s)} ,{k}_t'^{(s)} =1}}^{K}} \sum_{{k}_r ,{k}_t =1}^K 
     \text{Pr}( |r_{{k}_r,{k}_t}^{(s)}| > |r_{{k}_r'^{(s)},{k}_t'^{(s)}}^{(s)}|).
  \end{align}

\noindent
In order to derive an explicit expression for each probability term in (\ref{P_fail_eq_2}), we need to find the joint distributions of $r_{{k}_r,{k}_t}^{(s)}$ and $ r_{{k}_r'^{(s)},{k}_t'^{(s)}}^{(s)}$. For a given $\alpha$, by recalling (\ref{r_incorr}), (\ref{B_zero}) and (\ref{p_s}) we can see that $r_{{k}_r,{k}_t}^{(s)}$ and $ r_{{k}_r'^{(s)},{k}_t'^{(s)}}^{(s)}$ can form a jointly Gaussian distributed length-2 vector with mean

\begin{equation}
\label{mean_mat}
    \text{E}\left[ \! \! \! \! \begin{array}{c}
    r_{{k}_r,{k}_t}^{(s)} \\
    r_{{k}_r'^{(s)},{k}_t'^{(s)}}^{(s)}
    \end{array} \! \! \! \!  \right] \!= \! \left[  \! \! \! \begin{array}{c}
    \alpha \sqrt{P_T} \boldsymbol{b}_{{k}_r}^T \boldsymbol{b}_{k_r'^{(s)}} \boldsymbol{b}_{k_t'^{(s)}}^T \boldsymbol{b}_{{k}_t} x   \\
    \alpha \sqrt{P_T} x
    \end{array} \! \! \!  \right]\! \triangleq \! \left[ \! \!  \! \begin{array}{c}
    a \\
    b
    \end{array} \! \! \!  \right].
\end{equation}    

\noindent
and covariance matrix

\begin{equation}
\label{covar_mat}
    \text{Cov} \left[ \! \begin{array}{c}
    r_{{k}_r,{k}_t}^{(s)} \\
    r_{{k}_r'^{(s)},{k}_t'^{(s)}}^{(s)}
    \end{array} \! \right] = \! \left[  \! \begin{array}{cc}
    N_0 & \Sigma \\
	\Sigma & N_0   
    \end{array} \right] 
\end{equation}    

\noindent
where

    \begin{align}
    \label{covar}
    \Sigma \;=\; & \text{E}[(r_{{k}_r,{k}_t}^{(s)}- \text{E}[r_{{k}_r,{k}_t}^{(s)}])^H(r_{{k}_r'^{(s)},{k}_t'^{(s)}}^{(s)}-\text{E}[r_{{k}_r'^{(s)},{k}_t'^{(s)}}^{(s)}])] \nonumber \\    
    =\;& \text{E}[(\boldsymbol{b}_{{k}_r}^T\boldsymbol{Q}^{(s)}\boldsymbol{b}_{{k}_t})^H (\boldsymbol{b}_{{k}_r'^{(s)}}^T\boldsymbol{Q}^{(s)}\boldsymbol{b}_{{k}_t'^{(s)}}) ]  \nonumber \\
=\;& \text{E}[\boldsymbol{b}_{{k}_t}^T(\boldsymbol{Q}^{(s)})^H\boldsymbol{b}_{{k}_r} \boldsymbol{b}_{{k}_r'^{(s)}}^T\boldsymbol{Q}^{(s)}\boldsymbol{b}_{{k}_t'^{(s)}}^T ]  \nonumber \\    
    =\;& N_0 (\boldsymbol{b}_{{k}_r}^T \boldsymbol{b}_{k_r'^{(s)}} \boldsymbol{b}_{k_t'^{(s)}}^T \boldsymbol{b}_{{k}_t}). 
    \end{align}

\noindent
Then following equation (4B.21) of \cite{proakis}, we have

   \begin{align}
    \label{Q_m}
     \!\!\! \text{Pr}   \Big(  | r_{{k}_r,{k}_t}^{(s)}| >&  |r_{{k}_r'^{(s)},{k}_t'^{(s)}}^{(s)}|\Big| |\alpha| \Big)  = \nonumber \\ 
      \!\!&\!\!Q_1{(A,B)} - \frac{1}{2}I_0(AB)\text{exp}{(-\frac{1}{2}(A^2+ B^2))}
    \end{align}
 
\noindent 
where $Q_1$ is the first order generalized Marcum Q function, $I_0 $ is the 0th order modified Bessel function of the first kind, $A=|a|/\sqrt{N_0-\Sigma}$ and $B=|b|/\sqrt{N_0-\Sigma}$. By further assuming that the magnitude of the channel coefficient, $|\alpha|$, follows distribution $f_{|\alpha|}(x)=\frac{x}{\text{Var}[\alpha]} \text{exp}({-\frac{x}{2\text{Var}[\alpha]}})$, we can write each additive term in (\ref{Q_m}) as \cite{unified}

    \begin{align}
    \label{Q_m_rayleigh}
     \text{Pr}(|&r_{{k}_r,{k}_t}^{(s)}| > |r_{{k}_r'^{(s)},{k}_t'^{(s)}}^{(s)}|  )\nonumber \\ 
     &= \int_0^\infty  \text{Pr}\Big(| r^{(s)}_{{k}_r,{k}_t}| > |r_{{k}_r'^{(s)},{k}_t'^{(s)}}^{(s)}| \Big| |\alpha| \Big) f_{|\alpha|}(x) \mathrm{d} x \nonumber \\  
     &=   \frac{1}{2}-\frac{ \bar{B}^2-\bar{A}^2 }{ 4 \sqrt{1+\bar{B}^2+\bar{A}^2 + (\frac{\bar{B}^2}{2}-\frac{\bar{A}^2}{2})^2 } } 
     \end{align}
 
\noindent 
where $\bar{A}\!=\text{E}[A]=\text{Var}[\alpha]\sqrt{P_T} \boldsymbol{b}_{{k}_r}^T \boldsymbol{b}_{k_r'^{(s)}} \boldsymbol{b}_{k_t'^{(s)}}^T \boldsymbol{b}_{{k}_t} |x| /\sqrt{N_0-\Sigma}$ and $\bar{B}=\text{E}[B]=\text{Var}[\alpha] \sqrt{P_T}|x|/\sqrt{N_0-\Sigma}$. By substituting (\ref{Q_m_rayleigh}) into (\ref{P_fail_eq_2}) we have

    \begin{align}
     \label{p_fail_d}
     \!\!\!\!{P_{fail}}(s\Big|&s-1) \leq \frac{1}{K^2} {\sum_{{{k}_r'^{(s)} ,{k}_t'^{(s)} =1}}^{K}} {\sum_{{{k}_r ,{k}_t =1}}^{K}} \nonumber \\      
      &\bigg[ \frac{1}{2}-\frac{ \bar{B}^2-\bar{A}^2 }{ 4 \sqrt{1+\bar{B}^2+\bar{A}^2 + (\frac{\bar{B}^2}{2}-\frac{\bar{A}^2}{2})^2 } }  \bigg].
    \end{align}

\noindent
Finally the PCEF can be upper bounded by substituting (\ref{p_fail_d}) into (\ref{fail_cond_full}) such that

    \begin{align}
    \label{p_fail_final}
     P_{fail}(S)  \leq  & \frac{S}{K^2} {\sum_{{{k}_r' ,{k}_t' =1}}^{K}} {\sum_{{{k}_r ,{k}_t =1}}^{K}} \nonumber \\      
      & \bigg[ \frac{1}{2}-\frac{ \bar{B}^2-\bar{A}^2 }{ 4 \sqrt{1+\bar{B}^2+\bar{A}^2 + (\frac{\bar{B}^2}{2}-\frac{\bar{A}^2}{2})^2 } }  \bigg].
    \end{align}
    
\section{Numerical Results}

    \begin{figure}[!t]
    \centering
    \includegraphics[width=3.44in]{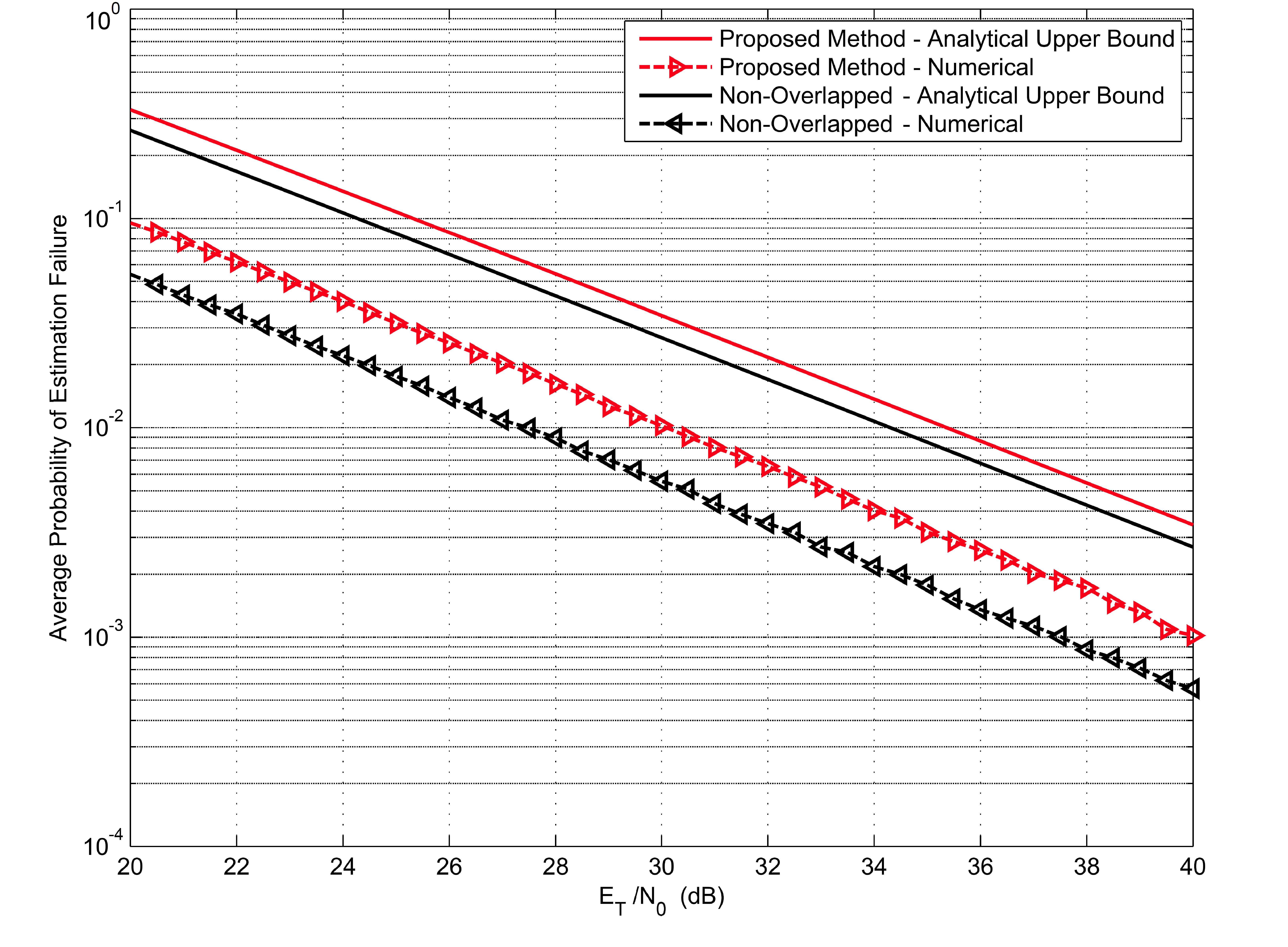}
    \caption{Comparison of the probability of failed channel estimation between the proposed algorithm and \cite{rheath}'s non-overlapped algorithm.}
    \label{performance}
    \end{figure}

We now provide some numerical examples to verify the performance of our proposed algorithm. We consider a single-path MMW system with $N = 27$ antennas at both the transmitter and receiver. The fading coefficient is assumed to follow a complex Gaussian distribution with zero mean and variance $N^2$. We set $K = 3$ for both our algorithm and the algorithm in \cite{rheath}. Thus, $K^2 = 9$ time slots are required in each stage of \cite{rheath}. However, in our proposed algorithm, only $M^2 = 4$ time slots are involved in in each stage. This significantly reduces the total time slots required for channel estimation. Power allocation among the $S$ stages is applied to both algorithms as (\ref{p_s}).

\figurename \ref{performance} shows the PCEF as described in (\ref{fail_cond}), where the total energy required in the overall channel estimation process is calculated by $E_T=M^2\sum_{s=1}^Sp_s$. We can see that, to achieve the same PCEF as the algorithm in \cite{rheath}, our proposed algorithm requires 2.5dB more energy in order to reduce the channel estimation time by $225\%$. The gap between the simulation results and derived upper bound is similar to that derived in \cite{rheath}. This gap originates from the relaxations in (\ref{fail_cond_full}) and (\ref{P_fail_eq_2}).

\begin{figure}[!t]
    \centering
    \includegraphics[width=3.4in]{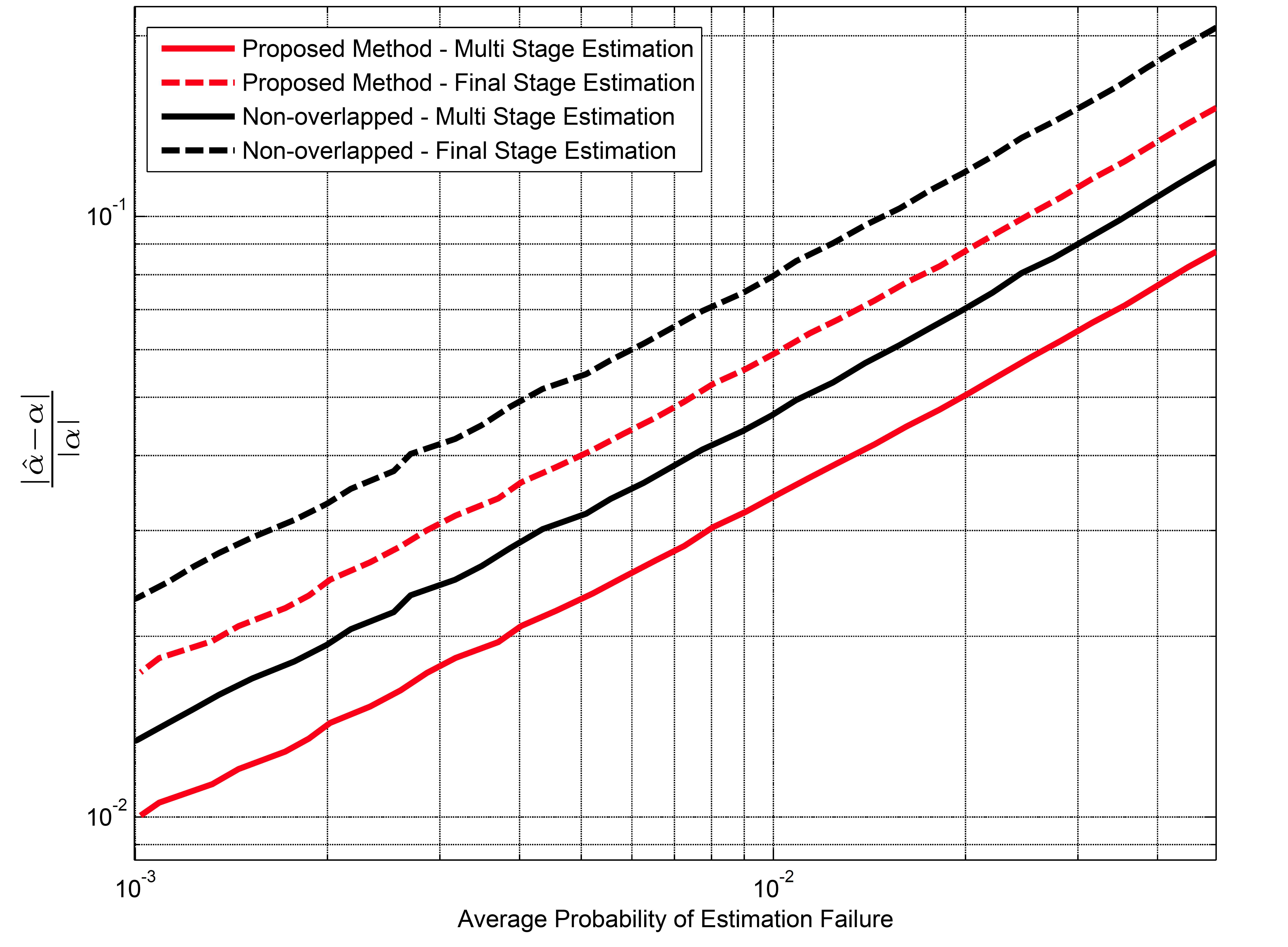}
    \caption{Comparison of relative error in fading coefficient estimation between the non-overlapped algorithm from \cite{rheath} and the proposed algorithm.}
    \label{time}
    \end{figure}

\figurename \ref{time} shows the relative estimation error of the fading coefficient $\alpha$, i.e., $\frac{|\hat{\alpha}- \alpha |}{|\alpha|}$. As can be seen, the proposed algorithm performs better than that in \cite{rheath} at the same PCEF as more energy is required in the former. \figurename \ref{time} also shows that by combining the selected measurements from all stages with MMSE estimator, we significantly improve the estimation accuracy for both algorithms. 

\section{Conclusion}

In this paper we have proposed a new channel estimation algorithm for MMW channels based on a novel overlapped beamform pattern design. In general, our proposed algorithm can hasten the channel estimation process by a factor of $\frac{K^2}{\text{log}_2^2(K+1)}$ when compared to the existing method developed in \cite{rheath} with the same value of $K$. For example, with $K=3$ the proposed algorithm reduces the number of measurement time slots by $225\%$ and will achieve the same PCEF as that in \cite{rheath} for a cost of 2.5dB more energy. For channels with rapidly changing channel information, this cost can be justified in order to improve the estimation speed.

\bibliographystyle{IEEEtran}
\bibliography{IEEEabrv,ICC2015}

\end{document}